%% file: slaw-igsp-hep-v3.tex
\documentclass[11pt,twoside]{gsphep}

\usepackage{hyperref}  

\usepackage{mathenv} 

\usepackage{calligra,pbsi}
\usepackage[T1]{fontenc}

\input somedef

\newcommand{\rd}{\mathrm{d}} 
\newcommand{\ri}{\mathrm{i}} 

\newcommand{\beq}{\begin{equation}}
\newcommand{\eeq}{\end{equation}}
\newcommand{\beqa}{\begin{eqnarray}}
 \newcommand{\eeqa}{\end{eqnarray}}
 \newcommand{\nn}{\nonumber}
\newcommand{\half}{\frac{1}{2}}

\newcommand{\uind}[2]{^{#1_1 \, ... \, #1_{#2}} }
\newcommand{\lind}[2]{_{#1_1 \, ... \, #1_{#2}} }

\newcommand{\co}{\circ}

\newcommand{\pbr}[2]{ \{ \hspace*{-2.6pt} [ #1 , #2\hspace*{1.4 pt} ] 
\hspace*{-2.6pt} \} }

\newcommand{\we}{\wedge}

\newcommand{\xx}[1]{\raisebox{1pt}{$\stackrel{#1}{X}$}}

\newcommand{\ff}[1]{\raisebox{1pt}{$\stackrel{#1}{F}$}}

\newcommand{\der}{\partial}

\newcommand{\inn}{\hspace*{2pt}\raisebox{-1pt}{\rule{6pt}{.3pt}\hspace*
{0pt}\rule{.3pt}{8pt}\hspace*{3pt}}}

\newcommand{\bd}{\mbox{\bf d}}

\newcommand{\ka}{\varkappa}

\newcommand{\Psib}{\overline{\Psi}}
\newcommand{\Phib}{\overline{\Phi}}

\newcommand{\what}[1]{\widehat{#1}}

   \newcommand{\betab}{\bar{\beta}}
 \newcommand{\gammab}{\bar{\gamma}} 
 
\newcommand{\omegab}{\overline{\omega}}

\newcommand{\bx}{{\mathbf{x}}}

\newcommand{\BPsi}{{\bf \Psi}} 
   
\newcommand{\BS}{{\bf S}}

\begin{document}

\thispagestyle{plain}

\title{Ehrenfest Theorem in precanonical quantization}

\author{I.V.~~Kanatchikov}

\date{}

\maketitle


\begin{abstract}
We discuss the precanonical quantization of fields which is based on the 
De Donder--Weyl (DW) Hamiltonian formulation and treats the space and 
time variables on an equal footing. Classical field equations in DW Hamiltonian 
form are derived as the equations for the expectation values of 
precanonical quantum operators. This field-theoretic generalization of 
the Ehrenfest theorem demonstrates the consistency 
of three aspects of precanonical field quantization: 
(i) the precanonical 
representation of operators in terms of the Clifford (Dirac) 
algebra valued partial differential operators, 
(ii) the Dirac-like precanonical generalization of the Schr\"odinger equation 
without the distinguished time dimension, and 
(iii) the definition of the scalar product for  calculation of 
expectation values of operators using the 
precanonical wave functions. 
\end{abstract}

\label{first}

{\footnotesize
\tableofcontents
}
\newpage

  {\footnotesize \bsifamily 
I am very honoured to contribute a paper to the volume 
 dedicated to Professor Jan S\l awianowski. I deeply appreciate his encouraging 
support during  my hard years in Warsaw in the second half of the 1990s. 
Some aspects of the Ehrenfest theorem in (what I later called) 
 precanonical quantization  
 of fields
were discussed with him at his Laboratory of Analytical Mechanics 
and Field Theory already around 1997. Moreover, one of my earlier 
attempts to understand 
a covariant field quantization leading 
in the classical limit to the 
generalized Hamilton-Jacobi 
theories in the calculus of variations  \cite{kastrup,rund} 
was inspired by the geometric discussion of the van Vleck determinant 
in S\l awianowski's  monumental book on the geometry of phase space 
\cite{slaw}. 
 } 

\section{Introduction}

The canonical Hamiltonian formalism in field theory 
is not the only possible extension of the Hamiltonian 
formalism from mechanics to field theories 
described by multiple integral variational problems 
(see e.g. \cite{kastrup,rund}). 
Moreover, the alternative 
extensions, such as the De Donder--Weyl (DW) 
 theory \cite{dedonder,weyl35}, actually do not need 
to distinguish a time dimension and, therefore, are not restricted to 
the globally hyperbolic space-times.  It is natural to ask if 
the alternative Hamiltonian formulations can lead to a certain 
reformulation of the quantization procedure in field theory, 
which would be more general than the canonical quantization. 
Though the DW theory has been known in the calculus of variations since the 1930s, 
 it is 
 the lack of a suitable generalization of the Poisson bracket to this framework 
 which 
 made it impossible to use 
 for field quantization. 
When such a generalization was found in 1993 \cite{my93,mybrackets2,mybrackets1}, it has paved the way to the 
 approach to 
 field quantization 
based on the DW theory, which I later called 
{\em precanonical quantization}. The term reflects 
 the nature of mathematical structures of the DW theory, which are 
 in a sense intermediate 
between the Lagrangian formalism 
and the canonical Hamiltonian formalism. 


The Ehrenfest theorem initially has been playing an important heuristic role 
in developing a field quantization based on the DW 
Hamiltonian formulation in field theory. However, the importance of this 
role is probably not obvious from the papers which I have published 
at different stages of the development of the theory \cite{myIJTP98a,myprecanonical,myprecanonical1,geomq}. In this paper 
I would like to present 
 a more systematic treatment 
of the Ehrenfest theorem 
in the 
quantum theory of fields which is based on precanonical quantization. 
 A more naive treatment, which is found in my earlier papers, is now 
 improved by a proper definition of the scalar product of Clifford-valued 
 precanonical 
 wave functions and a modified notion of 
 self-adjoint operators 
 with respect to  this scalar product, which comply with the fact that a quantum 
 formalism 
 resulting from precanonical 
 quantization is essentially the one with an indefinite metric Hilbert 
 space. 
 
 Note that the ability of precanonical quantization to reproduce 
the correct classical field equations on the average can be considered as 
a test of precanonical representation of operators, 
the precanonical analogue of the Schr\"odinger equation and 
the prescription for the calculation of expectation values 
of operators using the Clifford algebra valued precanonical 
wave functions. 

We proceed as follows. 
In Section 2 we discuss 
 the precanonical quantization starting from the 
outline of the DW Hamilonian formulation and the Poisson-Gerstenhaber  
brackets of differential forms, which generalize the Poisson brackets 
to the DW theory. The quantization based on these brackets is outlined in 
Section 2.3. In Section 2.4 we briefly discuss a connection between the 
precanonical field quantization and the functional Schr\"odinger 
representation in QFT. Different aspects of the Ehrenfest theorem 
in the context of precanonical field quantization are discussed in Sections 3--5. 
We consider the Ehrenfest theorem in the case of interacting 
scalar fields in flat space-time in Section 3,  pure Yang-Mills theory 
in Section 4, and the scalar fields in curved space-time in Section 5. 
The latter consideration allows us to identify the connection term 
in the curved space-time generalization of the precanonical Schr\"odinger 
equation with the spin-connection. 
The concluding remarks are found in Section 6. 

\section{Precanonical Field Quantization}

Let us first outline the basic elements of precanonical quantization. 
Instead of using the canonical Hamiltonian formalism, which requires 
a decomposition into the space and time, we start from the De Donder--Weyl 
extension of the Hamiltonian formulation of the Euler-Lagrange equations 
to field theory \cite{kastrup,rund}, 
where no distinction between the space and time variables is required. 

\subsection{De Donder--Weyl Hamiltonian Formulation}

Let us consider a  field theory  given by a Lagrangian density  
$L = L(y^a, y^a_\mu, x^\nu)$, 
which is a function of the space-time variables $x^\mu$, 
field variables $y^a$ and  the coordinates of 
their first space-time derivatives (first jets) $y^a_\mu$, 
such that on a specific field configuration 
$y^a=y^a(x)$, $y^a_\mu = \der_\mu y^a(x)$.  
 We  can  define  new Hamiltonian-like 
variables without  the distinction between the space and time variables: 
the {\em polymomenta} 
\beq 
 p^\mu_a := \frac{\der L}{\der y^a_\mu} 
 \eeq
and the {\em DW Hamiltonian function} 
\beq
 H(y^a, p^\mu_a, x^\mu) := y^a_\mu(y,p) p^\mu_a - L . 
\eeq
Then, if the DW Legendre transformation $(y^a, y^a_\mu) \rightarrow (y^a,p^\mu_a)$ 
is regular, i.e., 
\beq 
\det ||\der^2L/\der y^a_\mu \der y^b_\nu ||\neq 0 
\eeq 
the Euler-Lagrange field equation can be written in 
 {\em DW Hamiltonian form} 
$$ 
\der_\mu y^a (x) =
                            \frac{\der H}{\der p^\mu_a} , \quad 
\der_\mu p^\mu_a (x) = 
                                    - \frac{\der H}{\der y^a}   .
 \refstepcounter{equation}                                  
 \eqno {  (\theequation a,b) } \label{dw}
$$ 
In what follows we denote $\frac{\der}{\der y^a}$ as $\der_a$.

Note that it is also possible to construct 
an analogue of the Hamilton-Jacobi (HJ) theory 
associated with the DW Hamiltonian formulation. The corresponding DWHJ equation 
 \cite{kastrup,rund,weyl35}  
\beq \label{dwhj}
\der_\mu S^\mu + H \left( y^a, p^\mu_a = 
\der_a S^\mu, x^\mu\right) = 0 
\eeq 
defines the solutions of field equations in terms of the wave fronts corresponding to the eikonal functions $S^\mu(y^a,x^\mu)$ on the finite dimensional analogue of the configuration space, 
 i.e. the space of field variables $y^a$ and space-time variables $x^\mu$. 
The very existence of such a 
Hamilton-Jacobi theory on the finite dimensional space of $y^a$ and $x^\mu$  
rises the question about the existence of a formulation of 
quantum field theory in terms of the wave functions 
on this space, which leads to the DWHJ equation in the classical limit.

\subsubsection{Example:  Classical Interacting Scalar Fields} 

In the case of the theory  of interacting 
scalar fields $y^a$ with the Lagrangian 
\beq \label{Lscalar}
L = \frac12 \der_\mu y^a \der^\mu y_a - V (y) 
\eeq
where $V(y)$ includes both 
the mass terms like $\frac12 \frac{m^2}{\hbar^2}y^2$ 
 and the interactions, 
we obtain $p_a^\mu = \der^\mu y_a$ and 
\beq \label{hdw-scalar}
H=\frac12 p^\mu_a p_\mu^a + V(y). 
\eeq 
The DW Hamiltonian equations obtained from (\ref{dw}) 
\beq \label{dwscalar}
\der_\mu p^\mu_a = - \der_a V, \quad \der_\mu y^a = p_\mu^a  
\eeq
are just the first order form of the coupled nonlinear Klein-Gordon equations  
for scalar fields $y^a=y^a(x)$. 

The DWHJ equation (\ref{dwhj}) for interacting scalar fields 
takes the form of a partial differential equation 
\beq \label{dwhjscalar}
\der_\mu S^\mu + \frac{1}{2} \frac{\der S^\mu}{\der y^a} \frac{\der S_\mu}{\der y_a} + V(y)=0 
\eeq
where $S^\mu (y^a,x^\mu)$ are eikonal functions on the 
finite dimensional covariant configuration space. 
 By treating the space $\bx$ and 
time $t := x^0$ variables differently and constructing a functional 
$$\BS ([y^a(\bx)],t):= \int\! \rd\bx\ S^0(y^a=y^a(\bx),\bx,t) $$ 
 we can show \cite{my-pla2001} that, 
 as a consequence of the DWHJ equation (\ref{dwhjscalar}),  
the functional $\BS$ obeys the standard Hamilton-Jacobi equation in functional 
derivatives, which is familiar from the canonical Hamiltonian formalism  
 \beqa
\der_t\BS + \int\!dy\ \Big( 
\frac12 \frac{\delta\BS}{\delta y^a(\bx)}\frac{\delta\BS}{\delta y_a(\bx)} 
+ \frac12 \big(\nabla y^a(\bx)\big)^2 + V\big(y(\bx)\big) \Big) 
=0 . \nn  
 \eeqa
This is one of the examples of how the DW (precanonical) 
Hamiltonian structures 
{\em pre}cede the canonical ones.

\subsection{Poisson Brackets in DW Hamiltonian Formulation}

Quantization based on the DW Hamiltonian-like framework requires 
a suitable generalization of\ Poisson brackets. We found a generalization 
of the geometric construction of Poisson brackets in analytical mechanics 
(see e.g. \cite{slaw}) 
to the DW Hamiltonian framework, where it is based on a higher degree 
generalization of the symplectic structure to the extended 
{\em polymomentum phase space} 
of variables $z^M:=(y^a,p_a^\mu, x^\mu)$. Namely, this generalization 
is given by the {\em polysymplectic form}\footnote{This object can be defined 
as a representative of a certain equivalence class of forms, see \cite{mybrackets2}. 
For the related discussions see also  \cite{forger,deleon-book,roy,sardan}.} 
\cite{my93,mybrackets2}
\beq \label{ps}
\Omega = \rd p^\mu_a \we \rd y^a\we \varpi_\mu 
\eeq 
where $\varpi_\mu := \der_\mu \inn \varpi$ and $\varpi := \rd x^1\we..\we \rd x^n$. 
Thus, in field theory on $n$-dimensional space-time a generalization of the 
symplectic form is a form of degree $(n+1)$. The particular form of (\ref{ps}) 
follows from the Poincar\'e-Cartan (PC) form corresponding to the DW theory 
\cite{gimmsy} 
and the geometric representation of solutions of classical field equations 
in terms of multivector fields on the polymomentum phase space 
(see \cite{my93,mybrackets2} for details). 
Namely, the DW Hamiltonian equations can be represented 
as the equations of the integral surfaces of 
$n$-multivector fields $\xx{n}$, such as \cite{my93,mybrackets2} 
\beq \label{mh}
\xx{n}\inn\,\Omega = (-)^n \rd H . 
\eeq  
Thinking about the introduction of a Poisson bracket, we conclude that 
the map between $0$-forms and $n$-multivectors in (\ref{mh}) 
should be generalized to include the horizontal (semi-basic) 
forms of other degrees 
\beq \label{mf}
\xx{n-p}\inn\,\Omega = \rd \ff{p} , \quad p=0,1,...,(n-1) 
\eeq 
where 
$\ff{p}:= \frac{1}{p!}F_{\mu_1 ...\mu_p} (y^a,p^\nu_a,x^\nu)\,\rd x^{\mu_1}\we...\we \rd x^{\mu_1}. $
This map is also suggested by the {\em polysymplectomorphism} 
symmetry introduced in \cite{my93} in terms of the Lie derivatives 
with respect to the multivector fields. 
Note that the map in (\ref{mf}) exists only for a special class of forms called {\em Hamiltonian forms} in \cite{my93,mybrackets2} (see also \cite{mybrackets1} 
for an explicit formula for the Hamiltonian forms) 
and it maps those forms to the equivalence classes of multivector fields modulo the annihilators of\ $\Omega$: $\xx{p}\inn\,\Omega = 0, \quad p=2,...,n.$  

The above constructions lead to the following formula for the Poisson bracket of two Hamiltonian forms $\ff{p}_1$ and $\ff{q}_2$ 
\beq \label{pbra} 
\pbr{\ff{p}_1}{\ff{q}_2} = (-)^{(n-p)} \xx{n-p}_1\inn\, \rd \ff{q}_2 
\eeq
which gives rise to the graded Lie algebra structure on Hamiltonian forms, where the grade of a $p$-form with respect to the 
bracket operation is $(n-p-1)$. It is easy to see that the bracket of $p$ and $q$ forms is a Hamiltonian form of degree $(p+q-n+1)$. 

If we want a true Poisson bracket, we also need the bracket to obey 
an analogue of the Leibniz rule. From the definition of Hamiltonian forms in (\ref{mf}) it follows that Hamiltonian $p$-form is poly-linear 
of degree $(n-p)$ in polymomenta \cite{mybrackets1}.  Therefore, the exterior product of two Hamiltonian forms is not a Hamiltonian form in general. Nevertheless, we found the product operation  with respect to which the space of Hamiltonian forms is closed. It is called the {\em co-exterior product}\cite{mybrackets1} and denoted 
as $\bullet$ 
\beq \label{coext}
\ff{p}\bullet \ff{q}:= *^{-1} (*\ff{p}\we*\ff{q}) 
\eeq
where $*$ is the Hodge duality operator on the space-time. This product requires only a volume $n$-form on the space-time for its definition \cite{geomq}. 

Thus we see that a $p$-form has the grade $(n-p)$ with respect to the $\bullet$-product, which is different by one from its degree with respect to the bracket operation $\pbr{\,\cdot\, }{\cdot }$.  
We can also check that the bracket in (\ref{pbra}) 
is a graded derivation with respect to the co-exterior product, 
i.e. the graded Leibniz rule is fulfilled by the 
graded Lie bracket with respect to the graded commutative product $\bullet$. Therefore, the space of Hamiltonian forms with the operations $\pbr{\,\cdot\, }{\cdot }$ and $\bullet$ is the 
{\em Gerstenhaber algebra } \cite{mybrackets2,mybrackets1}. This structure generalizes the Poisson algebra structure to field theory within the 
DW Hamiltonian formulation. In this formulation the dynamical variables 
 are represented by the Hamiltonian forms on the polymomentum phase space. 

A connection between  the Poisson-Gerstenhaber brackets on forms in the DW theory 
and the standard Poisson brackets in the canonical Hamiltonian 
formalism, which are defined on the  functionals of field configurations 
in the canonical phase space, has been 
discussed in \cite{mybrackets2,helein,vey}.   

The bracket defined in (\ref{pbra}) allows us to calculate simple brackets between  the 
Hamiltonian 
forms constructed from the field and polymomenta variables, which will generalize the canonical 
brackets, viz. 
$$
\pbr{p_a^\mu\varpi_\mu}{y^b}
= 
\delta^b_a , \;
\pbr{p_a^\mu\varpi_\mu}{y^b\varpi_\nu}
=
\delta^b_a\varpi_\nu, \;
\pbr{p_a^\mu}{y^b\varpi_\nu}
= 
\delta^b_a\delta^\mu_\nu . 
\refstepcounter{equation} 
\eqno {  (\theequation a,b,c) } \label{fundbr}
$$

Moreover, the Poisson-Gerstenhaber bracket in (\ref{pbra}) allows us to write the equations of motion of 
Hamiltonian $(n-1)$-forms  $F:=F^\mu (y^a,p^\mu_a,x) \varpi_\mu$ in terms of the bracket with the DW Hamiltonian function $H$. In $n$-dimensional  Minkowski space
\beq \label{bracketform}
\bd\!\bullet\! F = 
                   (-1)^{n}\pbr{H}{F}  
+ \rd^h \!\bullet \! F 
\eeq
where $\bd\bullet$ denotes the {\em total co-exterior differential} 
of a $p$-form $\ff{p}$ 
\beq
\bd\!\bullet\! \ff{p}:= \frac{1}{(n-p)!} 
 \frac{\der}{\der z^M} F\uind{\mu}{n-p}\,\der_\mu z^M (x) 
\rd x^\mu\bullet \varpi\lind{\mu}{n-p} 
\eeq 
 $\varpi\lind{\mu}{n-p} := \der\lind{\mu}{n-p}\inn \varpi$, 
and $\rd^h$ is the {\em horizontal co-exterior differential} 
\beq
\rd^h\!\bullet \! \ff{p}:= \frac{1}{(n-p)!}\der_\mu F\uind{\mu}{n-p}   
\rd x^\mu\bullet \varpi\lind{\mu}{n-p}. 
\eeq

By substituting  the  $(n-1)$-form variables from 
the fundamental brackets (\ref{fundbr}) into (\ref{bracketform}) 
we reproduce the DW Hamiltonian equations (\ref{dw}). 
Note that  equation (\ref{bracketform}) generalises 
the Poisson bracket form of the equations of motion 
of a function on the phase space $F(q,p,t)$ 
in mechanics: $\frac{\rd}{\rd t}F = \{H,F\} + \der_t F$. 


\subsection{Precanonical Quantization} 

Precanonical quantization is based on a generalization 
of the Dirac rule of canonical quantization, which relates the Poisson brackets 
with the commutators of quantum operators, to the Poisson-Gerstenhaber 
brackets in the DW theory 
\beq
[\hat{A}, \hat{B}] = -\ri\hbar \what{\pbr{A}{B}} . 
\eeq
The mathematical and physical reasons of why the Dirac 
quantization rule allows us to obtain 
a 
quantum description from the classical one, 
though not uniquely, is a separate great issue,  
which we have very little to say about. 
Here we take it as a technical postulate of quantum theory. 

Let us quantize the fundamental precanonical brackets in (\ref{fundbr})  (see \cite{myprecanonical,myprecanonical1}). In the $y$-representation, when $y^b$ are 
multiplicative operators, from quantization of (\ref{fundbr}a)  
we obviously obtain 
\beq
\what{p^\nu_a\varpi}_\nu = -\ri\hbar \der_a 
\eeq
i.e. a classical $(n-1)$-form is represented by 
a quantum operator of form degree $0$. 
This representation is also consistent with quantization of 
(\ref{fundbr}b), which, however,  does not  specify the operator of 
the form  $\what{\varpi}_\nu$. 
Quantization of (\ref{fundbr}c) leads to the commutator 
\beq \label{21}
[\hat{p}{}^\mu_a,y^b\what{\varpi}_\nu] = 
\hat{p}{}^\mu_a \co y^b\what{\varpi}_\nu - y^b\what{\varpi}_\nu\co \hat{p}{}^\mu_a 
= 
\ri\hbar \delta^b_a\delta^\mu_\nu 
\eeq 
where $\co$ denotes a composition law of operators. Therefore, 
$\hat{p}{}^\mu_a = \ri\hbar \hat{\epsilon}^\mu \otimes \der_a$ and  
\beq \label{22}
\hat{\epsilon}^\mu \co \what{\varpi}_\nu = \delta^\mu_\nu 
, \qquad
\hat{\epsilon}^\mu \co \what{\varpi}_\nu - \what{\varpi}_\nu \co \hat{\epsilon}^\mu = 0.
\eeq
It is easy to see that these relations can be fulfilled if 
$\hat{\epsilon}^\mu$ and $\what{\varpi}_\nu $ are represented by 
Dirac matrices and $\co$ is their symmetric product, i.e., 
\beq
\what{\varpi}_\nu = \frac{1}{\ka} \gamma_\nu 
, \qquad 
 \hat{\epsilon}^\mu = \ka \gamma^\mu 
\eeq
where $\frac{1}{\ka}$ is a small constant of the dimension of $(n-1)$-volume, 
which appears on the purely dimensional grounds. Therefore, the polymomenta 
are represented  by 
 the 
 Clifford algebra valued operators 
 \beq \label{p-op}
 \hat{p}{}^\mu_a = - \ri\hbar\ka \gamma^\mu \der_a . 
 \eeq
 

The bracket form of field equations in (\ref{bracketform}) allows us to guess 
the form of 
{\em precanonical Schr\"odinger equation } 
\beq \label{nse}
\ri\hbar\varkappa \gamma^\mu\der_\mu \Psi =\what{H} \Psi 
\eeq 
where the precanononical wave function $\Psi$ is a Clifford-valued 
wave function on the finite dimensional covariant configuration space: 
$\Psi (y^a,x^\mu)$. 
In the following sections we will see that this form of the 
Schr\"odinger equation is consistent with the Ehrenfest theorem. 

Note that the Dirac operator in the left hand side 
of (\ref{nse}) is a  quantum version of 
$(-)^{n-1} \rd\bullet$, which is generated by the (commutator related 
to) the bracket with $H$ 
in (\ref{bracketform}).  
Hence, 
we can identify the quantum operator of $\rd x^\mu\bullet \ $ with 
$(-)^{n-1} \ka \gamma^\mu $. 
This observation will be used later 
in the calculation in equation (\ref{47}).

\subsubsection{Example: Quantum Interacting Scalar Fields}

We can obtain an explicit expression of the operator of 
the DW Hamiltonian for the system of interacting scalar fields 
(\ref{hdw-scalar})  by calculating the bracket 
\beq
\pbr{p^\mu_a p_\mu^a}{y^b\varpi_\nu} = 2 p^b_\nu 
\eeq
and quantizing it using the already known representation 
of $\hat{p}{}^\mu_a$ and $\what{\varpi}_\nu$. 
The result is 
\cite{myprecanonical,myprecanonical1,myIJTP98a}
\beq \label{hscalar}
 \hat{H} = -\frac{1}{2}\ \hbar^2\ka^2 
  \frac{\der^2}{\der y^a \der y_a} + 
V(y). 
\eeq 

For the free scalar field $V(y) \sim m^2 y^2$, so that $\hat{H}$ represents 
a harmonic oscillator in the space of field variables $y$.  
This theory can be easily solved and the 
precanonical wave functions can be written down explicitly 
(see e.g. \cite{myprecanonical1,my-rev15}).

\subsection{Precanonical Quantization and Standard QFT}

The functional Schr\"odinger representation is one of the standard 
descrptions of quantum fields, though not the most widely used one. 
There is an excellent textbook by Hatfield \cite{hatfield}, 
which treats many standard aspects of QFT using the functional 
Schr\"odinger representation. In this picture the states of quantum fields 
are described by the Schr\"odinger  wave functionals $\BPsi ([y^a(\bx)], t)$, 
which are functionals of field configurations $y^a(\bx)$ at a given 
instant of time $t$ (we use the notation $x^\mu:= (\bx, t)$). 

It is natural to ask how this description is related to the 
description in terms of precanonical 
wave functions $\Psi(y^a, x^\mu)$. A comparison of the probabilistic 
interpretations of  the Schr\"odinger wave functional 
$\BPsi ([y^a(\bx)], t)$ (an amplitude of finding 
a field configuration $y^a(\bx)$ at the instant $t$) 
and the precanonical wave function $\Psi(y^a,x^\mu)$ 
(an amplitude of finding a value of the field $y^a$ at the space-time point $x^\mu$) suggests that the former can be represented as 
a combination of the latter taken along a specific configuration $y^a=y^a(\bx)$. 
This idea has been explored in several papers 
\cite{myIJTP98a,my-pla2001,my-atmp14,my-atmp15} 
and it has resulted in the following formula, which expresses the Schr\"odinger wave functional in terms of the 
Volterra's multidimensional  {\em product integral\/}\cite{productintegral,slavik}  
 of precanonical wave functions restricted to  the surface $\Sigma$ 
 in the space of $(y^a,x^\mu)$, which  
 represents the field configuration $y=y(\bx)$ at the instant of time $t$ 
\beq\label{schr-prod}
\BPsi ([y(\bx)],t) = \Tr \left \{\prod_\bx 
e^{-\ri y(\bx)\alpha^i\der_iy(\bx) \rd\bx} 
  \Psi_\Sigma (y(\bx), \bx, t)_{\mbox{\large $\rvert$} \scriptscriptstyle    
  \frac{1}{\ka} \beta \mapsto \rd\bx }
\right \} .  
\eeq 
Here the notation $\Psi_{\mbox{$\rvert$}\scriptscriptstyle  \frac{1}{\ka} \beta \mapsto \rd\bx }$ 
means that every $\beta/\ka$ in the expression of $\Psi$ is replaced by $\rd\bx$ 
before the product integral is evaluated. In \cite{my-atmp14,my-atmp15}  
it is shown that the canonical functional derivative Schr\"odinger equation for $\BPsi ([y(\bx)],t)$ 
can be derived from the precanonical Schr\"odinger equation (\ref{nse}) in the 
 vanishing $1/\ka$ limit 
 or, more precisely, in the singular 
 limit when $\beta\ka$  is mapped to $\delta^{n-1}(\mbox{\bf 0})$. 
Formula (\ref{schr-prod}) is a consequence of this derivation. 
In \cite{my-atmp15} it has been  explicitly demonstrated  how 
equation (\ref{schr-prod})  allows us to construct the well known expression of the 
vacuum state wave functional of  the free scalar field \cite{hatfield} 
from the ground state solution of the precanonical 
Schr\"odinger equation 
for the free scalar field. 
 
The conclusion from those considerations is that the standard QFT 
obtained from the canonical quantization is a limiting case corresponding 
to an infinitesimal $\frac{1}{\ka} \rightarrow 0$ of the description of 
quantum fields obtained from the precanonical quantization.

 % ...

\section{Ehrenfest Theorem}

There has been some uncertainty regarding the nature of the 
wave function in precanonical quantization. 
In my earlier papers \cite{myIJTP98a,myprecanonical,myprecanonical1} 
I was tending to assume that the precanonical wave function 
$\Psi (y,x)$ 
is spinor-valued rather than Clifford algebra valued. 
One of the reasons was that the 
analogue of the Ehrenfest theorem was most straightforwadly provable 
with the spinor-valued wave functions. Besides, the positive 
definiteness of $\Psib\gamma^0\Psi$ for Dirac spinors, and the 
corresponding conservation law, which was following from the 
Dirac-like precanonical Schr\"odin\-ger equation (\ref{nse}), 
seemed to be a guarantee that the theory does have a meaningful 
probabilistic interpretation, in spite of the fact that the prescription 
of the calculation of expectation values of operators was based essentially 
on the scalar $\Psib \Psi$, which is not positive definite and even not 
preserved under 
 the space-time translations. Such a dichotomy of 
inner products is typical for the theories 
with an indefinite metric Hilbert space. Thus the principal advantage 
of preferring the Dirac spinor wave functions over the Clifford algebra 
valued wave functions seems to disappear and we have to  take seriously 
into account the 
fact that the quantum formalism which 
 follows 
from precanonical 
quantization is the one with an indefinite metric Hilbert space. 

In a later work on the relation of precanonical wave functions  
with the Schr\"odinger wave functional 
 \cite{my-atmp14,my-atmp15} 
we have seen that the constructions most naturally 
work for matrix-valued (i.e.,  the space-time Clifford (Dirac) algebra valued) 
precanonical $\Psi$-s, rather than for spinor-valued ones,  
i.e. valued in the  minimal ideals of the Clifford algebra. 
  
The treatment of the Ehrenfest theorem in this paper is different from 
our previous more naive considerations in that the precanonical wave 
function is taken to be Clifford algebra valued, and the definitions 
of the scalar product and the notion of self-adjointness of operators 
is consistent with the constructions known from the theories of 
the indefinite metric Hilbert spaces, with $\beta=\gamma^0$ playing 
the role of the so-called $J$-metric \cite{azizov}. 

If the wave function is a spinor $\Psi$, its conjugate is 
$\Psib := \Psi^\dagger\beta$. However, for a general Clifford-valued 
wave function the conjugate one is defined as $\Psib:=\beta \Psi^\dagger\beta$. 
By taking the Hermite conjugate of the  
precanonical Schr\"odinger equation (\ref{nse}) 
and multiplying it from the left and right by $\beta$, and assuming that 
the operator $\hat{H}$ is generalized self-adjoint in the sense that 
$\beta \hat{H}{}^\dagger\beta = H$,  we can write the equation of $\Psib$ 
in the form 
\beq \label{nse-conj}
\ri\hbar\ka\der_\mu \Psib \gamma^\mu = - \hat{H}\Psib 
\eeq 
where we have also used the property $\beta\gamma^\dagger{}^\mu\beta= \gamma^\mu$. 

Now we can prove the conservation law 
\beq
\der_\mu \int\! \rd y\ \Tr \big(\Psib \gamma^\mu \Psi\big) = 0 
\eeq 
where $\rd y:=\prod_a \rd y^a$. 

Indeed (for simplicity, we assume henceforth  in calculations 
that  $\hbar =1, \ka = 1$) 
\begin{align} 
\begin{split}
\ri\der_\mu \int\! \rd y\ \Tr \big(\Psib \gamma^\mu \Psi\big)
=& 
\int\! \rd y\ \Tr \big(\ri\der_\mu \Psib \gamma^\mu \Psi + \Psib \gamma^\mu \ri\der_\mu \Psi\big) 
\nn \\ 
=& \int\! \rd y\ \Tr \big(-\hat{H} \Psib  \Psi + \Psib  \hat{H}\Psi\big) 
 =0. 
 \end{split}\label{conserv}
\end{align}

Similarly, we can obtain 
\begin{align}
  \label{aver-p}
  \begin{split} 
\ri\der_\mu \int\! \rd y\ \Tr \big(\Psib \gamma^\mu \der_a \Psi\big)
=& 
\int\! \rd y\ \Tr \big(\ri\der_\mu \Psib \gamma^\mu \der_a \Psi 
+ \Psib \gamma^\mu \ri\der_\mu \der_a \Psi\big) 
\nn \\ 
=& \int\! \rd y\ \Tr \big(-\hat{H} \Psib  \der_a \Psi 
+ \Psib  \der_a \co \hat{H}\Psi\big) 
\nn \\ 
=& \int\! \rd y\ \Tr \big(- \Psib  \hat{H} \co \der_a \Psi 
+ \Psib  \der_a \co \hat{H}\Psi\big) 
\nn \\ 
=& 
 \int\! \rd y\ \Tr \big(\Psib  (\der_a\hat{H})  \Psi\big)  = \langle \der_a\hat{H}\rangle .
 \end{split}
\end{align}
Taking into account the precanonical representation of the operator of polymomenta 
(\ref{p-op}) this result shows that the  second  DW Hamiltonian equation 
(\ref{dw}b) 
is fulfilled on the average 
\beq \label{avp}
\der_\mu \langle \hat{p}^\mu_a\rangle = - \langle \der_a\hat{H}\rangle  
\eeq 
if the following prescription for the calculation of expectation values 
of precanonical operators is adopted 
\beq \label{averO}
\langle \hat{O}\rangle (x) = 
\int\! \rd y\ \Tr \Big(\Psib (y,x) \hat{O}  \Psi (y,x) \Big) .
\eeq
Note that the right hand side of (\ref{avp}) can be understood as follows: 
\beq
- \langle \der_a\hat{H}\rangle =  \langle [\hat{H},\der_a]\rangle 
=\langle [\hat{H}, \frac{\ri}{\hbar}\what{p_a^\nu\varpi_\nu} ] \rangle
= \langle \what{\pbr{H}{p_a^\nu\varpi_\nu}} \rangle .
\eeq

Next, let us consider 
\begin{align} \label{aver-y}
\der_\mu \langle y^a \rangle
&= \int\! \rd y \ \Tr \big(\der_\mu \Psib y^a \Psi +\Psib y^a \der_\mu\Psi \big) .
\end{align}
By multiplying the precanonical Schr\"odinger equation 
(\ref{nse}) and its conjugate (\ref{nse-conj}) 
by $\gamma^\mu$ we can write 
\beq \label{e1}
\ri\der_\mu \Psi = \gamma_\mu \hat{H}\Psi - \ri\gamma_{\mu\nu}\der^\nu \Psi 
, \qquad 
\ri\der_\mu \Psib = -\hat{H}\Psib \gamma_\mu + \ri\der^\nu \Psib  \gamma_{\mu\nu} .
\eeq
By substituting (\ref{e1}) 
into (\ref{aver-y}) we obtain
\begin{align} \label{e3}  
\begin{split}
\ri \der_\mu \langle y^a \rangle 
&= \int\! \rd y\ \Tr \Big( \big(-\hat{H}\Psib \gamma_\mu + \ri\der^\nu \Psib  \gamma_{\mu\nu}\big)y^a \Psi 
\\ 
&\hspace{63pt}+ \Psib y^a \big(\gamma_\mu \hat{H}\Psi - \ri\gamma_{\mu\nu}\der^\nu \Psi 
\big) \Big) \\  
&= \int\! \rd y\ \Tr \Big( \Psib \big( [y^a\gamma_\mu,\hat{H}] - \ri y^a \gamma_{\mu\nu} \stackrel{\leftrightarrow}{\der{}^\nu} \big)\Psi \Big ) 
\end{split}
\end{align}
where $a\stackrel{\leftrightarrow}{\der_\mu}{} b := a\der_\mu b - (\der_\mu a) b$. 

While the first term in (\ref{e3}) 
reproduces the statement of the Ehrenfest theorem for the 
first DW Hamiltonian equation in (\ref{dw}a), the nature of the second term 
is not clear. In fact, equations (\ref{e1})
are formal and their use 
should 
 take into account the integrability condition 
$\der_{[\mu}\der_{\nu ]}\Psi = 0$, which leads to a rather complicated system of additional equations. For this reason the use of equations (\ref{e1}) 
to prove the Ehrenfest theorem, in the way it is done in (\ref{e3}), 
does not appear to be justified. 

In order to prove the Ehrenfest theorem for the first DW Hamiltonian equation in (\ref{dw}a) by exploiting the same mechanism as in (\ref{aver-p}), 
let us use the fact that, according to the precanonical 
fundamental bracket in (\ref{fundbr}c), 
the variable (precanonically) conjugate to $p_a^\mu$ is an $(n-1)$-form 
$y^a\varpi_\nu$ ,  for which  equation (\ref{dw}a) can be rewritten as 
\beq \label{dwy1}
\der^\mu( y^a\varpi_\mu) = \frac{\der H}{\der p^\mu_a}\varpi_\mu = 
p_a^\mu\varpi_\mu 
\eeq
where in the last equality we use the expression of the DW Hamiltonian 
for the interacting scalar fields, see (\ref{hdw-scalar}). 
For the expectation value of the operator 
$\what{y^a\varpi_\nu} = \frac{1}{\ka}y^a\gamma_\mu$ we obtain 
\beqa
\ri\der^\mu\langle \what{y^a\varpi_\mu} \rangle 
&=& \ri\der_\mu\! \int\! \rd y\ \Tr \big(\Psib \gamma^\mu y^a \Psi\big) = 
\ri \!\int\! \rd y\ \Tr \big(\der_\mu \Psib \gamma^\mu y^a \Psi 
+ \Psib y^a \gamma^\mu \der_\mu  \Psi\big) 
\notag \\ 
&=& \int\! \rd y\ \Tr \big(-\hat{H} \Psib  y^a \Psi 
+ \Psib  y^a \hat{H}\Psi\big) 
 \label{pr2} \\
&=& \int\! \rd y\ \Tr \big( \Psib  [ y^a, \hat{H} ]  \Psi \big) 
  = 
 \int\! \rd y\ \Tr \big(\Psib  \der_a  \Psi\big)  = 
  \ri \langle \what{p_a^\mu\varpi_\mu}\rangle 
\nn
\eeqa
where 
 in the last line we use 
 the expression of the DW operator of interacting scalar fields  (\ref{hscalar}).

Thus, we have shown in (\ref{pr2}) that 
the first DW Hamiltonian equation in (\ref{dw}a) written in the form (\ref{dwy1}) 
is satisfied on the average as the equation for the expectation values of the 
corresponding operators. Together with equation (\ref{avp}) it proves the Ehrenfest theorem for the precanonically quantized system of interacting scalar fields 
in flat space-time: the classical  DW Hamiltonian equations  of this system are 
fulfilled by the expectation values of the corresponding precanonical  operators. 

However, there remains certain dissatisfaction due to the fact that we were able 
to prove the Ehrenfest theorem only for a specific form of the DW Hamiltonian 
equation: namely,  the one given by (\ref{dwy1}). 

Looking on the proofs in equations (\ref{aver-p}) and (\ref{pr2}), 
we see that 
the right hand sides  of the DW Hamiltonian equations are 
reproduced as expectation values of certain commutators 
with $\hat{H}$. It suggests that the Ehrenfest type 
statement is more naturally obtained for the Poisson bracket form of 
the DW Hamiltonian equations rather than for their naive form in (\ref{dw}). 

 Let us recall that in 
 the DW Hamiltonian theory we have shown that the DW Hamiltonian 
 equations in Minkowski space can be written in the form (cf. (\ref{bracketform})) 
\begin{align} \label{dwp}
\bd \bullet p^\mu_a\varpi_\mu 
=& 
(-)^n \pbr{H}{p^\mu_a\varpi_\mu} \\
\label{dwy}
\bd \bullet y^a \varpi_\mu 
=& 
(-)^n \pbr{H}{y^a\varpi_\mu} .
\end{align}
Equation (\ref{aver-p}) can be understood as tantamount to the following statement 
\beq
(-)^n\der_\mu \langle \what{\rd x^\mu\bullet}\co\what{p^\nu_a\varpi_\nu} \rangle 
= \langle \frac{\ri}{\hbar}[\what{H}, \what{p^\nu_a\varpi_\nu}]\rangle 
=  \langle  \what{\pbr{H}{p^\mu_a\varpi_\mu}}  \rangle 
\eeq
which is an Ehrenfestian version of (\ref{dwp}), 
 provided 
 $\what{\rd x^\mu\bullet}$ is identified with 
\linebreak $(-)^{n-1}\ka \gamma^\mu$. 
Note that the operator $\hat{\epsilon}^\mu$ in the representation of $\hat{p}^\mu_a$ 
in (\ref{22}) can be identified,
up to a sign factor,  with 
$\what{\rd x^\mu\bullet}$. An independent evidence of that could be 
in principle obtained also from the consideration of geometric quantization of 
Poisson-Gerstenhaber brackets in the DW Hamiltonian theory (see \cite{geomq}), 
given the fact that $\rd x^\mu\bullet$\ acts on forms similarly to the 
contraction with the multivector of degree $(n-1)$: $\epsilon^{\mu\mu_1...\mu_{n-1}}\der_{\mu_a}\we...\we\der_{\mu_{n-1}}$. 

Now, let us consider an Ehrenfestian version of equation (\ref{dwy}). The operator 
version of the r.h.s. of (\ref{dwy}):\  
$\bd \bullet y^a \varpi_\mu = \der_\nu (\rd x^\nu  \bullet y^a \varpi_\mu )$, 
can be written as $\der_\nu (\what{\rd x^\nu  \bullet}\co \what{y^a \varpi_\mu})$. 
Let us consider its average 
\begin{align} 
\begin{split} 
\der_\nu \langle \what{dx^\nu  \bullet}\co \what{y^a \varpi_\mu} \rangle 
=& \der_\nu\int\! \rd y\ \Tr \big(\Psib \what{dx^\nu  \bullet}\co \what{y^a \varpi_\mu} \Psi \big) 
\nn \\ 
&\hspace*{-55pt}=\;\int\! \rd y\ \Tr 
\big( \der_\nu\Psib \what{\rd x^\nu  \bullet}\co \what{y^a \varpi_\mu} \Psi 
+ \Psib \what{\rd x^\nu  \bullet}\co \what{y^a \varpi_\mu} \der_\nu\Psi \big) 
\nn \\ 
&\hspace*{-55pt}=\;  (-)^n \ri 
\int\! \rd y\ \Tr \big( \what{H}\Psib y^a \what{\varpi}_\mu 
-  \Psib y^a \what{\varpi}_\mu  \what{H}\Psi \big)
\nn \\ 
&\hspace*{-55pt}=\; (-)^n \ri  
\int\! \rd y\ \Tr \big(\Psib [\what{H},  y^a\what{\varpi}_\mu]\Psi \big) 
= (-)^n \langle \what{\pbr{H}{y^a\varpi_\mu}} \rangle 
\label{47}
\end{split}
\end{align} 
where in the third line we have used the property of the composition of 
operators $\what{\rd x^\mu\bullet}$ and $\what{\varpi}_\nu$: 
$\what{\rd x^\mu\bullet} \co \what{\varpi}_\nu 
- \what{\varpi}_\nu\co\what{\rd x^\mu\bullet} 
= 0,$ 
which results from quantization of one of the fundamental brackets 
in (\ref{21}), (\ref{22}). 
 Equation (\ref{47}) shows 
 that the bracket form of the second DW Hamiltonian equation 
(\ref{dwy})  is also fulfilled on the average.


\section{Ehrenfest Theorem in Pure Yang--Mills Theory} 

The Lagrangian density of pure Yang--Mills theory reads 
\beq
L= - \frac{1}{4} F_{a\mu\nu}F^{a\mu\nu} 
\eeq
where 
\beq
F^a_{\mu\nu} := \der_\mu A^a_\nu - \der_\nu A^a_\mu 
+ g C^a{}_{bc} A^b_\mu A^c_\nu 
\eeq 
$g$ is the Yang-Mills  self-coupling constant and 
$C_{abc}$ are totally antisymmetric structure constants which 
fulfill the Jacobi identity  
\beq
C^e{}_{ab}C^d{}_{ec} + C^e{}_{bc}C^d{}_{ea} + C^e{}_{ca}C^d{}_{eb}=0.
\eeq


The polymomenta and the DW Hamiltonian are given by 
\begin{align}
\pi_a^{\nu\mu} 
:=&\ \frac{\der L}{\der(\der_\mu A^a_\nu)} 
= -\der^\mu A_a^\nu + \der^\nu A_a^\mu - g C_a{}_{bc} A^b_\mu A^c_\nu 
=  F_a^{\nu\mu} \\
H
=&\
\pi_a^{\nu\mu}\der_\mu A^a_\nu - L 
 = 
-\frac{1}{4} \pi_{a\mu\nu} \pi^{a\mu\nu} 
+ \frac{g}{2} C^a{}_{bc}A^b_\mu A^c_\nu \pi_a^{\mu\nu} . 
\label{ym-dwh}
\end{align} 
The definition of polymomenta leads to the primary constraint 
(in the sense of the DW Hamiltonian theory\footnote{An extension of 
the Dirac's theory of constraints and the Dirac bracket 
to the DW Hamiltonian theory has been discussed in \cite{mydirac}.}) 
\beq \label{prim-ym}
\pi_a^{\mu\nu}+\pi_a^{\nu\mu} \approx 0  . 
\eeq
The Yang-Mills field equations in  DW Hamiltonian form read: 
\beqa \label{dw-ym-p}
\der_\mu \pi^{\nu\mu}_a &= -\frac{\der H}{\der A^a_\nu} 
 &= 
-g\, C_{abc}A^b_\mu\pi^{\nu\mu}_c  
\\
\der_{[\mu} A^a_{\nu]} &= \, \frac{\der H}{\der \pi^{\nu\mu}_a}  
 &= \frac{1}{2} \pi_{\mu\nu}^a - 
\frac{1}{2}g\, C^a_{bc}A^b_\mu A^c_\nu . \label{dw-ym-a}
\eeqa
The antisymmetrization in the left hand side of the 
second equation makes the  DW Hamiltonian equations 
consistent with the primary constraints. 

 Let us note that the 
related treatments of classical YM theory within 
the multisymplectic framework 
can be found 
in \cite{kondracki,marco,helein14}.
Precanonical quantization of YM theory, 
its connection with the functional Schr\"odinger representation, 
and a potential application to the mass 
gap problem have been discussed earlier 
in \cite{my-ym}.

Precanonical quantization leads to the representation of polymomenta as 
\beq
\hat{\pi}_a^{\mu\nu} 
= - \ri\hbar\varkappa \gamma^\nu  \der_{A^a_\mu} . 
\eeq 
The primary constraint (\ref{prim-ym}) is taken into account as 
the constraint on the physical quantum states 
\beq
\hat{\pi}_a^{(\nu\mu)}\left |\Psi\right >{}^{\rm phys} =0 
\eeq
whence it follows 
$\langle \hat{\pi}_a^{(\nu\mu)} \rangle{}^{\rm phys} =0$. 
From (\ref{ym-dwh}) we obtain the DW Hamiltonian operator 
\beq
\what{H} = 
 \frac{1}{2} \hbar^2\varkappa^2 \frac{\der}{\der A_a^\mu\der A^a_\mu } 
- \frac{1}{2} \ri g\hbar\varkappa  C^a{}_{bc}A^b_\mu A^c_\nu 
\gamma^\nu \frac{\der}{\der A^a_\mu } \; . 
\eeq
Note that in quantum YM theory the DW Hamiltonian operator is not scalar 
and the second term, which is responsible for self-interaction,  
is  Clifford algebra valued. 

The quantum states are represented by  
Clifford-valued wave functions $\Psi (A^\mu_{a}, x^\nu)$ with the scalar 
product given by 
\beq
\langle \Phi|\Psi\rangle = \int\![\rd A]\ \Tr\big(\Phib \Psi \big) 
\eeq
where the measure $[\rd A] = \prod_{a,\mu}\rd A^a_\mu$. 
The conservation law 
\beq
\der_\mu \int\! [\rd A]\ \Tr\Big(\Psib \gamma^\mu \Psi\Big) = 0
\eeq
follows from the precanonical Schr\"odinger equation (\ref{nse})  
and its conjugate (\ref{nse-conj}), 
and the fact that the DW Hamiltonian operator of pure YM system 
is generalized self-adjoint in the sense that 
$\hat{H} = \beta \hat{H}{}^\dagger\beta$, because 
$\beta\gamma^\mu{}^\dagger\beta= \gamma^\mu$.  

Now, a straightforward calculation yields 
\begin{align} 
\begin{split}
\der_\nu \langle \hat{\pi}{}^{\mu\nu}_a\rangle 
=& 
 -\ri\hbar\ka \der_\nu \int\! [\rd A]\ \Tr\Big(\Psib \gamma^\nu \der_{A^a_\mu} \Psi 
\Big) \nn \\ 
=&\int\! [\rd A]\ \Tr \Big( (\hat{H}\Psib) \der_{A^a_\mu} \Psi 
- \Psib \der_{A^a_\mu}\co \hat{H}\Psi\Big) = - \langle \der_{A^a_\mu}\hat{H} \rangle .
\end{split}
\end{align} 
Therefore, the first of the YM field equations in DW Hamiltonian form, 
equation (\ref{dw-ym-p}), 
is proven to be satisfied on the average.  

The validity of the Ehrenfest theorem for the second YM field equation (\ref{dw-ym-a}) 
can be proven similarly to the calculation in (\ref{47})
\newcommand{\alternative}{
\beq
+ - \der_{[\nu}A^a_{\mu]} = 
+- \bd\bullet (A^a_{[\mu}\varpi_{\nu]} = 
\frac{1}{2} \pi_{\mu\nu}^a - 
\frac{1}{2}g\, C^a_{bc}A^b_\mu A^c_\nu
\eeq
} 
\begin{align}
\begin{split}
 \der_{[\nu}A^a_{\mu]} 
 &= 
(-)^n \der_\alpha \langle ({A^a_{[\mu} \rd x^\alpha\bullet \co \varpi_{\nu]}})^{op}\rangle 
\nn \\ 
&=  (-)^n\der_\alpha \int\! [\rd A]\ \Tr\Big(\Psib 
{A^a_{[\mu} \what{\rd x^\alpha\bullet } \co \what{\varpi}_{\nu]}}\Psi \Big) 
\nn \\
&= \ri \int\! [\rd A]\ \Tr\Big(\hat{H}\Psib A^a_{[\mu}\what{\varpi}_{\nu]}\Psi - \Psib A^a_{[\mu}\what{\varpi}_{\nu]} 
\hat{H}\Psi \Big) 
\nn \\
&=  \ri \int\! [\rd A]\ \Tr\Big(\Psib 
[\hat{H},A^a_{[\mu}\hat{\varpi_{\nu]}}] \Psi \Big)  
\nn \\
&=  \int\! [\rd A]\ \Tr\Big(\Psib 
( \pbr{H}{A^a_{[\mu}\varpi_{\nu]}} )^{op} \Psi \Big )
=  \Big\langle \what{\frac{\der {H}}{\der \pi_a^{\mu\nu}}} \Big\rangle .
\end{split}
\end{align}

Thus, we have shown that the DW Hamiltonian form of YM field equation 
arises as the equation for the expectation values of precanonically quantized operators.


\section{Ehrenfest Theorem in Curved Space-Time} 

Let us consider interacting scalar fields on curved space-time background $g^{\mu\nu}(x)$. 
The dynamics is given by the Lagrangian density 
\beq
{\frak L} = \frac{1}{2} \sqrt{g}g^{\mu\nu} \der_\mu y^a \der_\nu y_a - \sqrt{g} V(y) 
\eeq 
where $g:=|\det g_{\mu\nu}|$,  and the designation of the parametric dependence 
from $x$-s is omitted here and in what follows.  
In this case the polymomenta 
\beq 
\frak{p}^\mu_a = \frac{\der \frak{L}}{\der \der_\mu y^a} = \sqrt{g}g^{\mu\nu} \der_\mu y_a 
\eeq 
the DW Hamiltonian density 
\beq
{\frak H} = \sqrt{g} H = \frac{1}{2\sqrt{g}} g_{\mu\nu} \frak{p}_a^\mu \frak{p}^{a\nu} + \sqrt{g} V(y) 
\eeq
and the polysymplectic structure 
\beq \label{pscurv}
\Omega  = \rd \frak{p}^\mu_a\we \rd\phi^a\we \varpi_\mu 
\eeq
are densities of the weight $+1$, 
which 
parametrically depend on the space-time coordinates $x$. 
Note that in our notation 
the  differentials $\rd$ in (\ref{pscurv}) 
do not act on $x$-s, as they are "vertical" (for the mathematical 
details of the definition of this notion, see \cite{mybrackets2}).


The DW Hamiltonian equations of the system of scalar fields given by ${\frak L}$ 
read 
\beq \label{dw-curved}
\der_\mu \frak{p}^\mu_a (x) = - \frac{\der \frak{H}}{\der y^a}, \qquad 
\der_\mu y^a(x)  =  \frac{\der \frak{H}}{\der \frak{p}^\mu_a} 
\eeq 
where $\der_\mu$ acts both on the parametric dependence on $x$ via $g^{\mu\nu}(x)$ 
and the dependence on $x$ due to the pull back to a specific section 
in the polymomentum phase space of variables $(\frak{p}^\mu_a, y^a)$, 
which represents a solution of classical field equations.  
Note that we could obtain the same equations by applying the usual rules of 
covariantization to the DW equations in flat space-time. 

The Poisson bracket operation defined by the weight $+1$ density valued polysymplectic structure (\ref{pscurv}) has a density weight $-1$, so that, for example,  
\beq
\pbr{\frak{p}^\mu_a(x)}{y^b\varpi_\nu } = \delta^b_a\delta^\mu_\nu .
\eeq


The Dirac quantization rule  in curved space-time is also modified  
to make sure that density valued quantities are quantized as density 
valued operators of the same weight 
\beq
[\hat{A}, \hat{B}] = - \ri\hbar\sqrt{g} \what{\pbr{A}{B}} . 
\eeq
It leads to the following representations 
\beq
\hat{\frak{p}}{}^\mu_a = -\ri\hbar\ka\sqrt{g} \gamma^\mu \der_a , 
\qquad 
\what{H} = -\frac{1}{2}\hbar^2\ka^2\der_a\der^a + V(y) 
\eeq
where the $x$-dependent $\gamma$-matrices are introduced such that 
$\gamma^\mu\gamma^\mu+ \gamma^\mu\gamma^\mu = 2g^{\mu\nu}$. 
Note that the operator of the DW Hamiltonian does not contain $x$-dependent quantities. 


The curved space-time version of 
 the precanonical Schr\"odinger equation (\ref{nse}) 
takes the form 
\beq \label{pse-curved}
\ri\hbar\ka\gamma^\mu(x)\nabla_\mu\Psi = \what{H}\Psi 
\eeq
where $\nabla_\mu := \der_\mu + \omega_\mu (x) $ is a covariant derivative of 
Clifford algebra valued wave functions. 
Let us see if the requirement that 
the Ehrenfest theorem 
extends also to the case of curved space-time  
can help us to specify the connection term $\omega_\mu (x)$.  

Before we proceed, let us find the precanonical Schr\"odinger equation 
for the conjugate wave function  $\Psib := \betab\Psi^\dagger\betab$,  
where  
$\gammab^I$, $I=0,...,n-1$ 
denote the flat (tangent) space Dirac matrices, such that 
$\gammab^I\gammab^J + \gammab^J \gammab^I = 2\eta^{IJ}$,  $\eta^{IJ}$ is 
the Minkowski metric, and $\betab := \gammab{}^0$. 
If $\hat{H}$ is generalized self-adjoint: $\hat{H} = \betab\hat{H}{}^\dagger \betab$, 
by multiplying the Hermite conjugate of (\ref{pse-curved}) by $\betab$ from 
the left and right, and inserting $\betab{}^2 =1$ where needed, we obtain 
\beq
\ri\hbar\ka \Psib ( \stackrel{\leftarrow}{\der_\mu} + \omegab_\mu ) \gamma^\mu 
= - \hat{H} \Psib 
\eeq
where $\omegab_\mu := \betab\omega_\mu^\dagger\betab$ (not to be confused 
with $\varpi_\mu$ in (\ref{pscurv})!). 

Let us consider a conservation law, which would generalize (\ref{conserv}) 
to curved space-time 
\beqa
\ri\der_\mu \! \int\! \rd y\ \Tr \Big(\Psib \sqrt{g}\gamma^\mu \Psi \Big) 
&=& \ri \int\! \rd y\ \Tr \Big(\der_\mu \Psib\sqrt{g} \gamma^\mu \Psi 
+ \Psib\sqrt{g} \gamma^\mu \der_\mu\Psi  
\nn \\ 
&& \hspace*{130pt} + \Psib \der_\mu (\sqrt{g}\gamma^\mu) \Psi 
\Big)\vspace*{-10pt}  \nn \\ 
\vspace*{-0pt} && \\
&&\hspace*{-120pt}=\,\int\! \rd y\ \Tr \Big( 
\Psib \sqrt{g}\big( -\hat{H} - \ri\omegab_\mu\gamma^\mu \big)\Psi + 
\Psib \sqrt{g} \big( \hat{H} -\ri\gamma^\mu\omega_\mu \big)\Psi + 
\Psib \ri\der_\mu (\sqrt{g}\gamma^\mu) \Psi \Big) \nn \\ 
&&\hspace*{-120pt}=\,\int\! \rd y\ \Tr \Big( \Psib 
\ri \big(- \sqrt{g}\omegab_\mu\gamma^\mu - \sqrt{g}\gamma^\mu\omega_\mu 
+ \der_\mu (\sqrt{g}\gamma^\mu)\big) \Psi \Big).  \nn
\eeqa
Therefore, the covariant version of the conservation law  
(\ref{conserv}) 
\beqa
\der_\mu \! \int\! \rd y\ \Tr \Big(\Psib \sqrt{g}\gamma^\mu \Psi \Big) =0 
\nn
\eeqa
is fulfilled if the connection $\omega_\mu$ satisfies the equality 
\beq \label{theta-condition}
\sqrt{g}\omegab_\mu\gamma^\mu + \sqrt{g}\gamma^\mu\omega_\mu 
- \der_\mu (\sqrt{g}\gamma^\mu) = 0. 
\eeq

Furthermore, 
\begin{align} \label{curvav-p}
\begin{split}
\ri\der_\mu\! \int\! \rd y\ \Tr \Big(\Psib \sqrt{g}\gamma^\mu \der_a \Psi \Big)
=& 
\int\! \rd y\ \Tr \Big(\ri\der_\mu \Psib \sqrt{g}\gamma^\mu \der_a \Psi 
\nn \\ 
& \hspace*{-80pt} +\; \Psib \der_a\sqrt{g}\gamma^\mu \ri\der_\mu  \Psi  
+  \Psib \ri\der_\mu (\sqrt{g} \gamma^\mu )\Psi  \Big) . 
\end{split}
\end{align}
By substituting 
the precanonical Schr\"odinger equation 
in curved space-time and its conjugate we obtain 
in the r.h.s. of ({\ref{curvav-p}}) 
\begin{align} 
\begin{split}
& \int\! \rd y\ \Tr \Big (-\sqrt{g}\hat{H} \Psib\der_a\Psi 
-\ri \Psib \sqrt{g}\omegab_\mu\gamma^\mu \der_a \Psi  
\nn \\ 
& \hspace*{80pt}+ \;  \Psib  \der_a \co \sqrt{g} \big(\hat{H} 
-  \ri \gamma^\mu\omega_\mu\big) \Psi 
+  \ri \Psib \der_\mu (\sqrt{g}\gamma^\mu )\der_a\Psi \Big) .
\end{split}
\label{eq70}   
\end{align}
The terms with $\hat{H}$ yield  
\begin{align} 
\begin{split}
&\int\! \rd y\ \Tr \Big(- \Psib  \sqrt{g}\hat{H} \co \der_a \Psi 
+ \Psib  \der_a \co \sqrt{g}\hat{H}\Psi \Big) 
\nn \\
&\hspace*{120pt}= \int\! \rd y\ \Tr \Big(- \Psib  (\der_a\hat{\frak{H}})  \Psi\Big)  
= - \langle \der_a\hat{\frak{H}}\rangle . 
\end{split}
\end{align} 
Therefore,  the first DW Hamiltonian equation in 
(\ref{dw-curved}) is fulfilled on the average if 
the remaining three terms in (\ref{eq70}) 
\beq 
 \int\! \rd y\ \Tr \Big ( \Psib 
 \big ( - \sqrt{g}\omegab_\mu\gamma^\mu 
-   \sqrt{g} \gamma^\mu\omega_\mu   
 +   \der_\mu (\sqrt{g}\gamma^\mu) \big ) \der_a\Psi \Big ) 
\eeq
produce a vanishing result. 
This condition limits the choice of the  connection 
$\omega_\mu$ 
and  it coincides with (\ref{theta-condition}).



 \newcommand{\toresore}{ 
According to the quantization rule in curved space-time 
quantization of the fundamental bracket 
$\pbr{\frak{p}{}^\mu_a}{y^b\varpi_\nu} =\delta_a^b\delta^\mu_\nu$ 
leads to the commutator 
\beq
[\hat{\frak{p}}{}^\mu_a,  \what{y^b\varpi_\nu}]= -\ri\ hbar 
\sqrt{g}\delta_a^b\delta^\mu_\nu
\eeq 
.... whence $\what{dx^\mu\bullet} = \sqrt{g}\gamma^\mu$ 
....
\begin{align} 
\der_\nu\langle \what{dx^\mu\bullet}\co \hat{\varpi_\mu}y^a\rangle 
=& \int\! \rd y\ \Tr \Big ( \der_\nu\Psib \sqrt{g} 
\what{dx^\mu\bullet}\co \hat{\varpi_\mu} y^a \Psi + 
\Psib\what{dx^\mu\bullet}\co \hat{\varpi_\mu} y^a \der_\nu \Psi 
\nn \\ 
\vspace*{-10pt}+\vspace*{-10pt}& 
 \Psib \der_\nu \sqrt{g}\gamma^\nu\co \hat{\varpi_\mu} y^a \Psi 
\end{align}
   } 

Now, let us consider the covariant version of 
 equation (\ref{e3}) 
\beq \label{55}
\nabla_\mu (y\varpi^\mu) = \der_\mu (y^a \varpi^\mu) + \half y \der_\mu (\ln g) \varpi^\mu . 
\eeq
Let us see if we can obtain it on the average from the precanonical 
Schr\"odinger equation on curved space-time. By a straightforward calculation 
we obtain 
\beqa
&&\ri\der_\mu \langle \what{y^a \varpi^\mu} \rangle=
\ri\int\! \rd y\ \Tr \Big( \der_\mu \Psib y^a\gamma^\mu \Psi + \Psib y^a\gamma^\mu \der_\mu\Psi +\Psib y^a (\der_\mu\gamma^\mu ) \Psi \Big) \nn \\
&&\!\!\!= \int\! \rd y\ \Tr \Big(\Psib\big(-\stackrel{\leftarrow}{\hat{H}}
-\ri\omegab_\mu\gamma^\mu\big)  y^a \Psi 
+ \Psib y^a \big(\hat{H}-\ri\gamma^\mu\omega_\mu\big) \Psi 
+\Psib y^a (\ri\der_\mu\gamma^\mu)\Psi 
\Big) \nn \\
&&\!\!\!= \int\! \rd y\ \Tr \Big(\Psib [y^a,\hat{H}] \Psi 
+\ri\Psib\big(-\omegab_\mu\gamma^\mu - \gamma^\mu\omega_\mu + \der_\mu\gamma^\mu\big)y^a \Psi \Big) . 
\eeqa
Therefore,  equation (\ref{55}) 
and the second DW Hamiltonian equation in (\ref{dw-curved})  
are fulfilled on the average if the connection $\omega_\mu$ 
satisfies the condition 
\beq
\omegab_\mu\gamma^\mu + \gamma^\mu\omega_\mu  - \der_\mu\gamma^\mu 
= \half \der_\mu (\ln g) \gamma^\mu 
\eeq
which is 
again 
equivalent to the condition obtained in (\ref{theta-condition}).  

One can view equation ({\ref{theta-condition}) for 
 the connection term as a consequence of 
a covariant constancy of the curved space-time Dirac matrices $\gamma^\mu(x)$ 
or, equivalently, the vielbeins $e^\mu_I (x)$. This is what identifies 
the connection term $\omega_\mu$ in (\ref{pse-curved}) with 
the spin-connection: $\omega_\mu = \omega_\mu^{IJ}\gammab_{IJ} = -\omegab_\mu$ 
with real coefficients $\omega_\mu^{IJ}$. 
 As the Clifford-valued precanonical wave function can be also viewed as 
 a spinor field with two spinor indices originating from the indices of 
 $\gamma$-matrices, 
 the appearance of the spin connection in the Dirac operator 
 in (\ref{pse-curved}) is natural here 
 (see e.g. \cite{redkov} for a related discussion). 
 



\section{Conclusions} 

We have shown that in the context of precanonical quantization 
of fields the evolution (or rather, space-time variation) of 
 expectation values of fundamental operators 
is consistent with classical field equations in DW Hamiltonian form. 
This property can be considered as a consistency test of 
three different aspects of precanonical quantization playing together: 
the precanonical representation of quantum operators 
in terms of Clifford-valued operators, 
the precanonical Schr\"odinger equation in (\ref{nse}), 
and the scalar product  for the calculation of 
expectation values of operators using the 
Clifford-valued precanonical wave functions in (\ref{averO}).  

We have explicitly demonstrated  that the Ehrenfest theorem 
can be proven for the system 
of interacting scalar fields both in flat and curved space-time, 
and for precanonically quantized pure Yang-Mills theory. 
In curved space-time the consideration of the Ehrenfest theorem 
 leads to the condition on the admissible connection term 
 in the Dirac operator   
 in the precanonical Schr\"odinger equation,
which 
is compatible with the known properties of 
the spin-connection. 

In our recent papers we have considered an application of precanonical quantization 
to the problem of quantization of gravity both in metric \cite{pqg-metric} 
and in vielbein \cite{pqg-vielbein} variables. We hope that it will be 
possible to demonstrate that the Einstein equations are also 
satisfied on the average as a consequence of our precanonical Schr\"odinger 
equation for quantum gravity, the precanonical representation of quantum operators 
appearing in our formulation, and the definition of the analogue 
of the Hilbert space of the theory which, in vielbein formulation 
\cite{pqg-vielbein}, involves an operator-valued measure on the space 
of spin-connection coefficients. 





\addcontentsline{toc}{section}{References}

Igor Kanatchikov \\
School of\, Physics  $\&$ Astronomy \\
University of\, St Andrews \\
 North Haugh, St Andrews \\
KY 16 9SS, Scotland \\
{\it E-mail address}: {\tt kanattsi@gmail.com}\\
\hspace*{73pt}{\tt ik25@st-andrews.ac.uk} 
\label{last}
\end{document}

%% file: somedef.tex
\def\Tr{
{\rm Tr}}

\def\det{
{\rm det}}

\renewcommand{\i}{\mathrm{i}}

\theoremstyle{plain}